\documentclass[
    ,final            
  ]
  {aipproc}

\usepackage{amssymb}
\layoutstyle{8x11single}

\begin{document}

\title{Short Gamma-Ray Bursts from Binary Neutron Star Mergers}

\classification{98.70.Lt, 95.30.Lz, 95.30.Sf, 97.60.Jd}
\keywords      {Gamma-Ray bursts, Hydrodynamics, Relativity and
gravitation, Neutron stars}

\author{Roland Oechslin}{
  address={Max Planck-Institut f\"ur Astrophysik, Karl
Schwarzschild-Str. 1, 85741 Garching, Germany}
}

\author{Thomas Janka}{
}


\begin{abstract}
 We present the results from new relativistic hydrodynamic
 simulations of binary neutron star mergers using realistic non-zero
 temperature equations of state. We vary several unknown
 parameters in the system such as the neutron star (NS) masses, their spins
 and the nuclear equation of state. The results are then investigated
 with special focus on the post-merger torus-remnant
 system. Observational implications on the Gamma-ray burst (GRB) energetics are
 discussed and compared with recent observations. 
\end{abstract}

\maketitle


\section{Introduction}

Merger events of binary neutron stars (BNS) and neutron star - black
hole (NS+BH) binaries do not only belong to the
strongest known sources of gravitational wave (GW) radiation, they are
also widely favored as origin of the subclass of short, hard
Gamma-ray bursts (GRBs) \citep{blinnikov1984, paczynski1986, eichler1989,
paczynski1991, narayan1992}.
The recent first good localizations of short bursts by the {\em
Swift} and {\em Hete} satellites were interpreted as a possible
confirmation of this hypothesis \citep{fox2005,
hjorth2005a,hjorth2005b, lee2005, bloom2006}, because the bursts have
observational characteristics which are different from those of long
GRBs, but which are in agreement with expectations from compact object
mergers as e.g. the energetics and the association with old stellar populations.
The central engines of such bursts are still poorly understood
and observationally undetermined. But it seems unlikely that
the energies required for typical short GRBs are set free during
the dynamical phase of the merging of two NSs. Instead, the following secular
accretion phase in a postmerger system consisting of a central BH and
a surrounding torus seems to be
a much more promising source (see e.g.,
\citep{woosley1993b,ruffert1999,popham1999,rosswog2003b,lee2005b})
provided the torus is sufficiently
massive. Thermal energy release preferentially above the poles of the BH by the annihilation
of neutrino-antineutrino ($\nu\bar\nu$) pairs can lead to
collimated, highly relativistic jets of baryonic matter
\citep{aloy2005}. The gamma radiation is then produced in internal
shocks when different blobs of ultrarelativistic matter in the jet
collide with each other. When the jet hits the ambient interstellar
medium, the less energetic afterglow radiation is produced.

Torus mass as well as BH mass and rotation are thus crucial parameters
that determine the energy release from the merger event. Here, we
summarize results from our recent 3D relativistic hydrodamics simulations of
BNS mergers \citep{oechslin2006a, oechslin2006b} to investigate the
dependence of the torus and BH properties on the initial BNS
parameters like the nuclear EoS, the NS masses and mass ratio and the
NS spins. 

In the following, we define the torus by the condition
that the specific angular momentum of a fluid element in the torus has
to be larger than the innermost stable circular orbit of a Kerr black
hole with the mass and the rotation parameter of the remnant. The
latter in turn is defined as all the matter which is not in the torus.

\section{Results from Relativistic Hydrodynamics Simulations of BNS mergers}

Using our relativisic smooth particle hydrodynamics (SPH) code
\citep{oechslin2006b}, we have simulated a large set of BNS merger
models, varying the EoS, the initial neutron star mass ratio, the total mass and the NS spins.

We have used two realistic, non-zero temperature EoSs whose
supranuclear parts base upon two fundamentally different approaches to determine the nuclear
interaction. The EoS of \citet{shen1998,shen1998b} (``Shen-EoS'') bases
on the phenomenological relativistic mean field (RMF) approach, while
the EoS of \citet{lattimer1991} (``LS-EoS'') employs the
finite-temperature compressible liquid droplet model for the nucleus
\citep{LLPR}. In addition, we consider an ideal gas EoS whose adiabatic
index $\Gamma$ is adjusted to fit the Shen-EoS in the supranuclear regime.

Initial data is obtained using a relaxation technique which drives the
system into an equilibrium velocity field which is
given by the orbital separation, the NS spins $\Omega_1$ and
$\Omega_2$, respectively and the orbital
angular velocity $\Omega$. The latter is iteratively determined such
that an orbital equilibrium state between gravitational and centrifugal force
is established (see \cite{oechslin2006b} for an extensive
description). The NS spins are chosen either to be zero (irrotating NSs),
or to be equal (corotating NSs) or opposed (counterrotating NSs) to the
orbital spin. A few cases with non-aligned NS spins have also been
considered (see Tab. \ref{tab:initialmodels}).

In addition, we
assume the initial NSs to be cold and to be in neutrino-less
beta-equilibrium which determines the initial distribution of the internal energy $u$ and the
electron fraction $Y_e$ (In the ideal gas case, the polytropic
relation $u=K\rho^{\Gamma-1}$, with $K$ fitted to the Shen-EoS, is
employed to initialize the internal energy). In Tab. \ref{tab:initialmodels}, we
summarize the parameters of our models. The simulations are carried
out with typically 400'000 SPH particles from slightly outside the
innermost stable circular orbit, through the late inspiral, merging
and torus formation until either the collapse of the merger remnant
sets in, or a quasi-stationary torus-remnant state has formed.\\

\begin{table}
\caption{Setup parameters and characteristic data of our considered models. $M_1$ and $M_2$
denote the individual gravitational masses in isolation while $M_\mathrm{sum}=M_1+M_2$
stands for the sum of the two (Note that the total gravitational mass $M$ is slightly smaller than $M_\mathrm{sum}$ because $M$
also involves the negative gravitational binding energy between the
two stars.). $M_0$ is the total baryonic mass, whereas $q=M_1/M_2$ and
$q_{M}=M_{0,1}/M_{0,2}$ is the gravitational and baryonic mass ratio, respectively. `Shen' stands for the full, finite 
temperature Shen-EoS, `Ls' denotes the Lattimer-Swesty-EoS and `ideal
gas' means ideal gas EoS. The NS spin states are denoted as `irrot' for
irrotating,
`corot' for corotating, `counter' for counterrotating, `oppo' for
oppositely oriented spins and `tilted' for a spin orientation tilted
relative to the orbital spin. More specifically, we choose in model
S1414t1 ${\Omega}_1=0$ and ${\Omega}_2=0.041*(0,1,0)$ (in cartesian coordinates) which
corresponds to a spin period of the second NS of $\sim 1$~ms, while in
model S1414t2, ${\Omega}_1=0.01*(0,1,1)$ and
${\Omega}_2=0.01*(0,-1,1)$ is chosen.
$T_\mathrm{max}$ is the maximum temperature in MeV reached in the system
during the whole evolution. It is obtained by averaging the particle
temperatures on a grid of $1.5$km side length. $a_\mathrm{merging}=J_\mathrm{merging}/M_\mathrm{merging}^2$ is
the (total) spin parameter measured immediately after merger and
$M_\mathrm{torus}$ denotes the estimated (baryonic) torus mass. Note
that in model LS1216 the maximal temperature and the torus mass cannot
be determined because the merger remnant collapses immediatedy after merging.}

\label{tab:initialmodels}
\begin{tabular}{c|c|c|c|c|c|c|c|c c|c|c|c}
Model & $M_1$ & $M_2$ & $M_\mathrm{sum}$ & $M_0$ & $q$ & $q_{M}$ &EoS& Spin& &$T_\mathrm{max}$&$a_\mathrm{merging}$&$M_\mathrm{torus}$\\

\hline
S1414 & 1.4 & 1.4 & 2.8& 3.032 & 1.0 & 1.0 &Shen&irrot&$\times ~\times$&52&0.91&0.06 \\ 
S138142 & 1.38 & 1.42 &2.8& 3.032 & 0.97 & 0.97&Shen&irrot&$\times
~\times$&50&0.90&0.06 \\ 
S135145 & 1.35 & 1.45 &2.8& 3.034 & 0.93 & 0.93&Shen&irrot&$\times ~\times$&50&0.90&0.09\\ 
S1315 & 1.3 & 1.5 &2.8& 3.037 & 0.87 &0.86&Shen&irrot&$\times
~\times$&50&0.90&0.15 \\ 
S1216 & 1.2 & 1.6 &2.8& 3.039 & 0.75 &0.73&Shen&irrot&$\times ~\times$&54&0.89&0.23
\\ 
S1515 & 1.5 & 1.5 &3.0& 3.274 & 1.0 &1.0&Shen&irrot&$\times ~\times$&67&0.89&0.05 \\  
S1416 & 1.4 & 1.6 &3.0& 3.274 & 0.88 &0.86&Shen&irrot&$\times ~\times$&55&0.89&0.17\\ 
S1317 & 1.3 & 1.7 &3.0& 3.279  & 0.76 &0.75&Shen&irrot&$\times
~\times$&65&0.88&0.23 \\ 
S119181& 1.19 & 1.81 & 3.0 & 3.289 & 0.66 & 0.63 &Shen&irrot&$\times ~\times$&57&0.87&0.24\\
S107193& 1.07 & 1.93 & 3.0 & 3.306 & 0.55 & 0.52 &Shen&irrot&$\times ~\times$&70&0.84&0.26 \\
S1313 & 1.3 & 1.3 &2.6& 2.800 & 1.0 &1.0&Shen&irrot&$\times ~\times$&52&0.93&0.08\\ 
S1214 & 1.2 & 1.4 &2.6& 2.799 & 0.86 &0.85&Shen&irrot&$\times ~\times$&43&0.93&0.20\\ 
S1115 & 1.1 & 1.5 &2.6& 2.807 & 0.73 &0.71&Shen&irrot&$\times ~\times$&50&0.92&0.23\\ 
\hline
P1315 & 1.3 & 1.5 & 2.8& 3.064 & 0.87 & 0.86 & ideal gas&
irrot&$\times ~\times$&&0.90&0.15\\
\hline
LS1414 & 1.4 & 1.4 & 2.8& 3.077 & 1.0 & 1.0 &Ls&irrot&$\times
~\times$&175&0.86& 0.03\\ 
LS1216 & 1.2 & 1.6 &2.8& 3.087 & 0.75 &0.73&Ls&irrot&$\times ~\times$&&0.85&-\\ 
\hline
S1414co & 1.4 & 1.4 & 2.8& 3.032 & 1.0 & 1.0 &Shen&corot&$\uparrow ~\uparrow$&40&0.98&0.24\\ 
S1414ct & 1.4 & 1.4 & 2.8& 3.032 & 1.0 & 1.0 &Shen&counter&$\downarrow
~\downarrow$&74&0.85&0.04\\ 
S1414o & 1.4 & 1.4 & 2.8& 3.032 & 1.0 & 1.0 &Shen&oppo&$\uparrow
~\downarrow$&46&0.92&0.11 \\ 
S1414t1 & 1.4 & 1.4 & 2.8& 3.032 & 1.0 & 1.0 &Shen&tilted&$\leftarrow
~\times$&59&0.95&0.20\\ 
S1414t2 & 1.4 & 1.4 & 2.8& 3.032 & 1.0 & 1.0 &Shen&tilted&$\nearrow
~\nwarrow$&39&0.98&0.26\\ 
S1216co & 1.2 & 1.6 &2.8& 3.039 & 0.75 &0.73&Shen&corot&$\uparrow
~\uparrow$&48&0.98&0.30\\ 
S1216ct & 1.2 & 1.6 &2.8& 3.039 & 0.75 &0.73&Shen&counter&$\downarrow
~\downarrow$&73&0.84&0.20\\ 
\hline

\end{tabular}
\end{table}

\begin{figure}
  \begin{minipage}[t]{0.32\linewidth}
   \includegraphics[width=5cm]{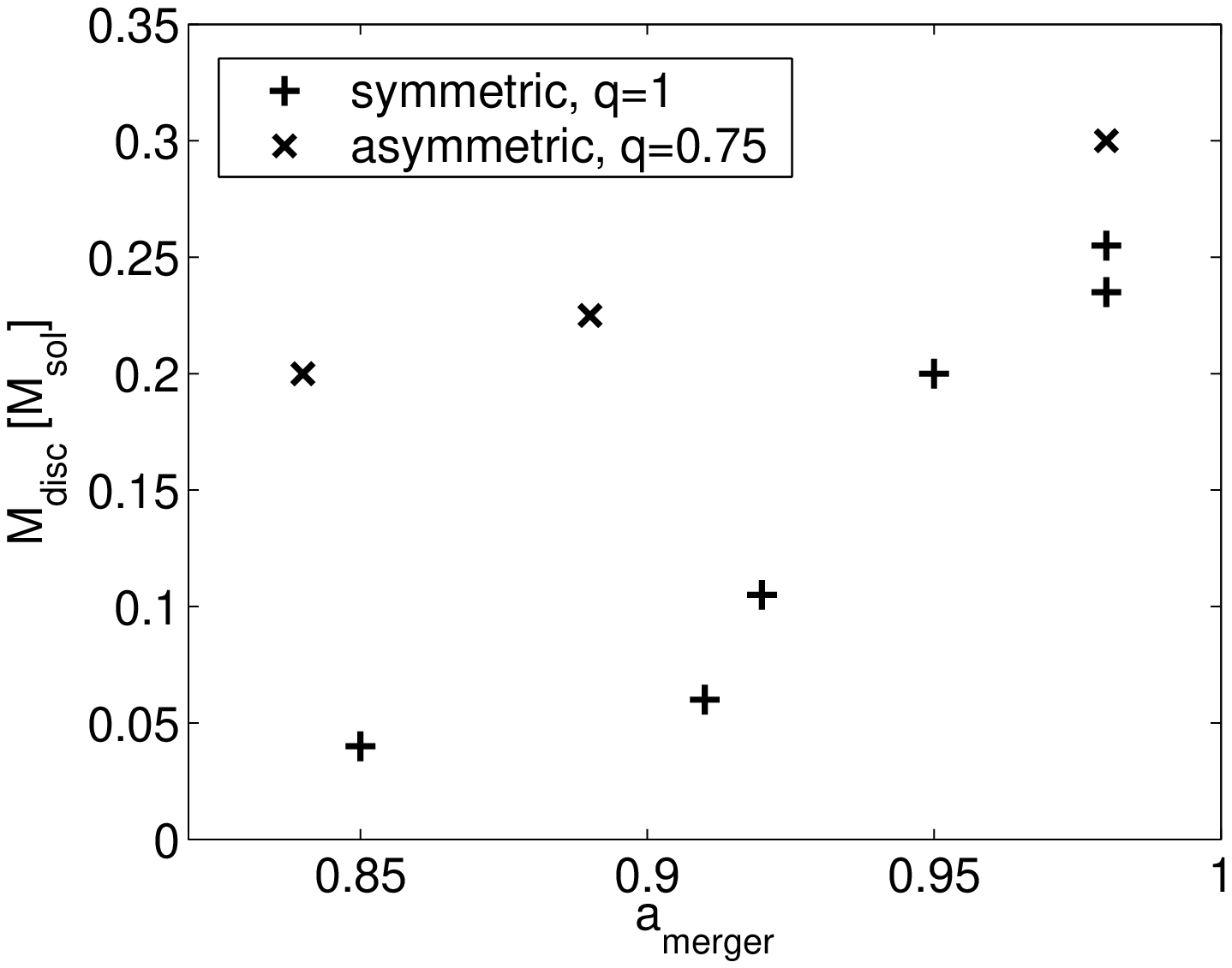}
	(a)
\end{minipage}
  \begin{minipage}[t]{0.32\linewidth}
   \includegraphics[width=5cm]{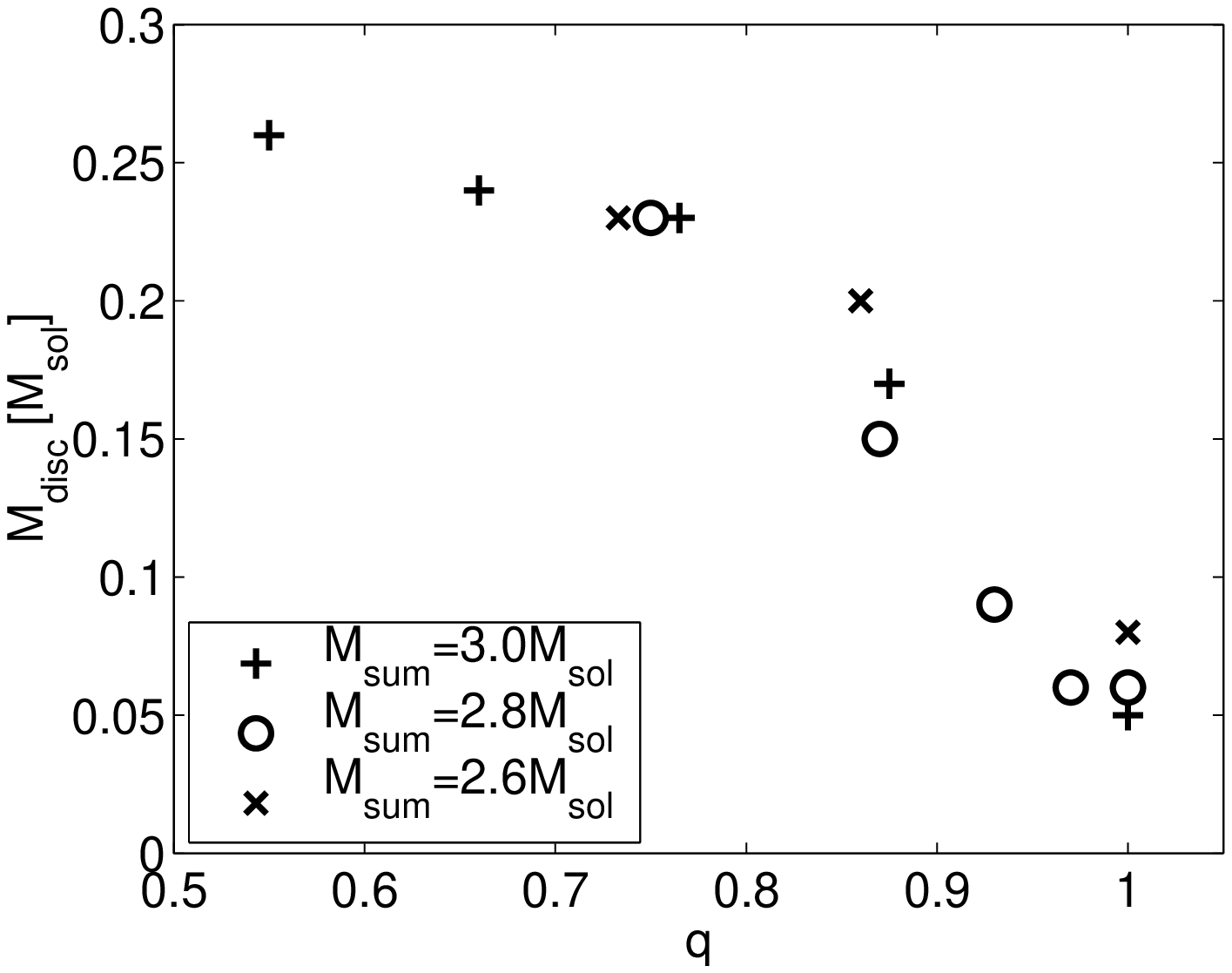}
	(b)
\end{minipage}
  \begin{minipage}[t]{0.32\linewidth}
   \includegraphics[width=5cm]{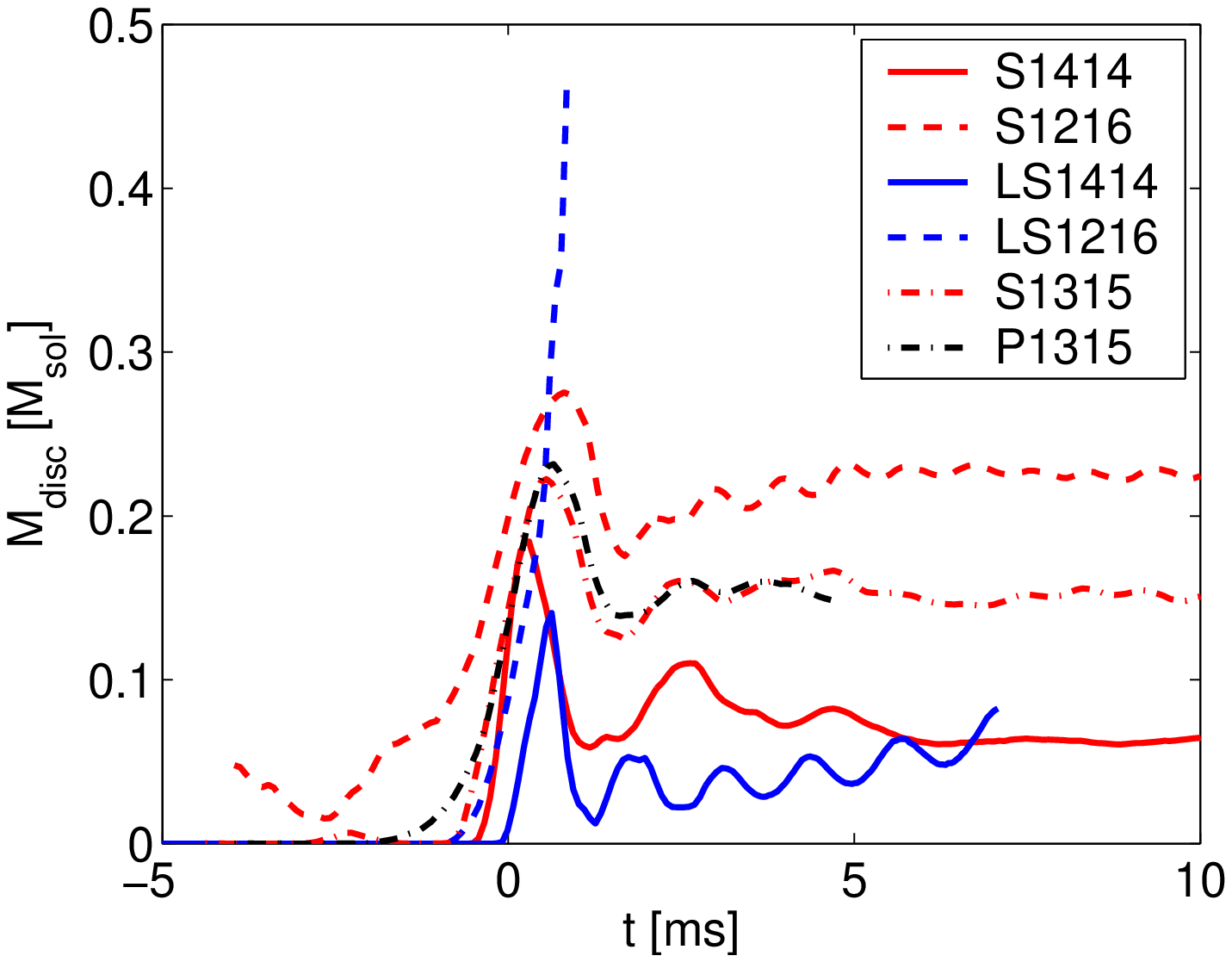}
	(c)
\end{minipage}
\caption{(a)~Torus masses versus total spin parameter
$a_\mathrm{merger}=J_\mathrm{merger}/M_\mathrm{merger}^2$ of the
system shortly after merger. Plotted are two series of models, with fixed EoS and
with mass ratio $q$=1 (`+'-signs) and $q$=0.75 (`x'-signs),
respectively.(b)~Torus masses versus mass ratio $q$. Three different
values of the total gravitational mass are chosen and marked with
`x'-sigs ($M_\mathrm{sum}=2.6M_\odot$), circles
($M_\mathrm{sum}=2.8M_\odot$) and `+'-signs
($M_\mathrm{sum}=3.0M_\odot$). These models use the Shen-EoS and are
initially irrotating.  (c)~Torus masses for different EoSs. The Shen-EoS
models are plotted in red, while the LS-EoS models are colored in blue and the ideal gas EoS model in
black. The more compact LS-EoS model LS1414 leads to a significantly
smaller torus mass then the corresponding Shen-EoS model S1414 shortly
after merger. During the subsequent evolution, angular momentum
transport by non-axisymmetric tidal interaction and numerical
viscosity increases the torus mass. Note that the peaks initially after merger are an
artifact of the non-linear feedback between torus mass, remnant mass and the angular
momentum of the last stable orbit in our torus-determination algorithm.
}
\label{fig:discmasses}
\end{figure}

\begin{figure}
  \begin{minipage}[t]{0.47\linewidth}
   \includegraphics[width=7.5cm]{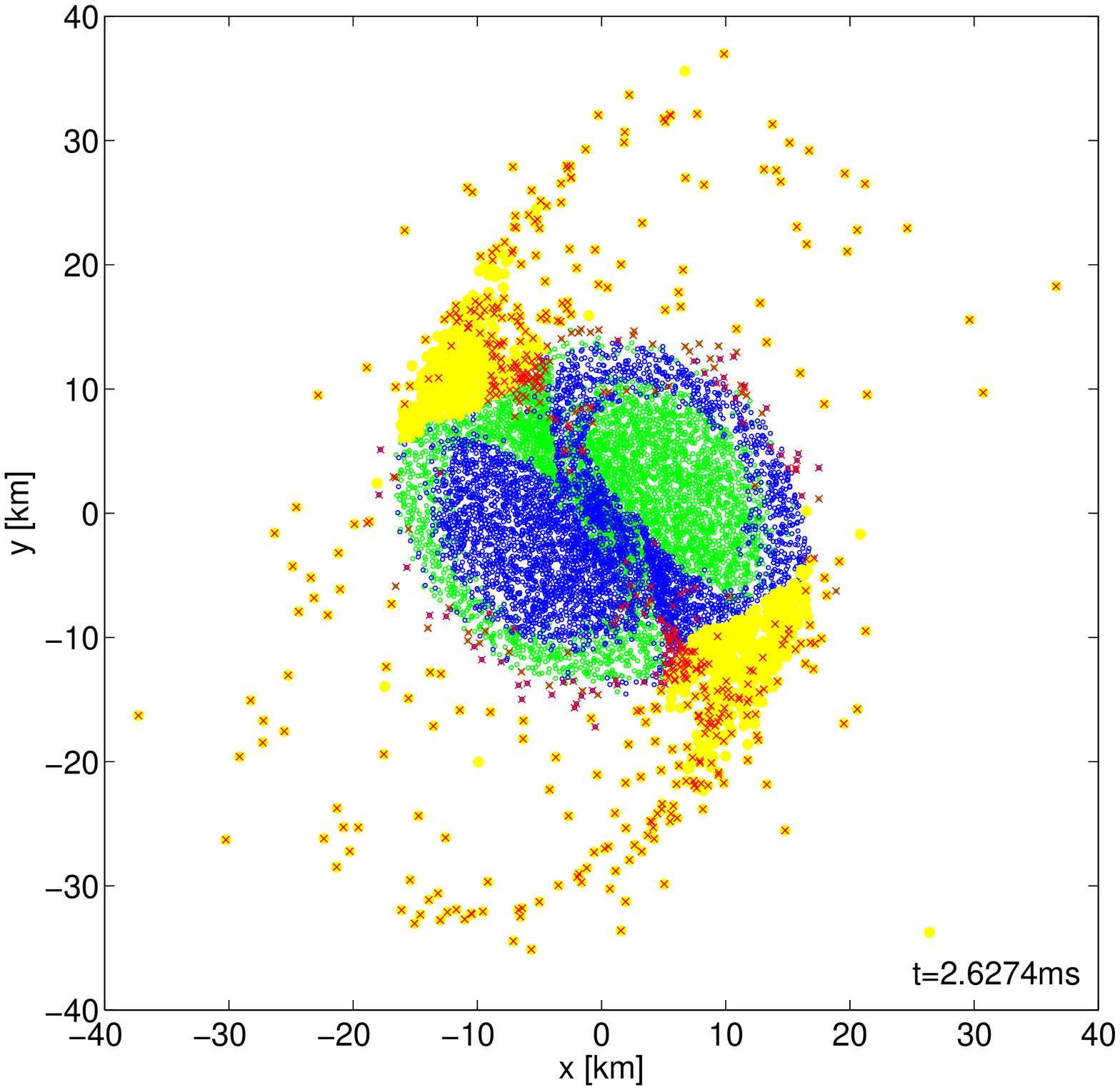}
	(a)

   \includegraphics[width=7.5cm]{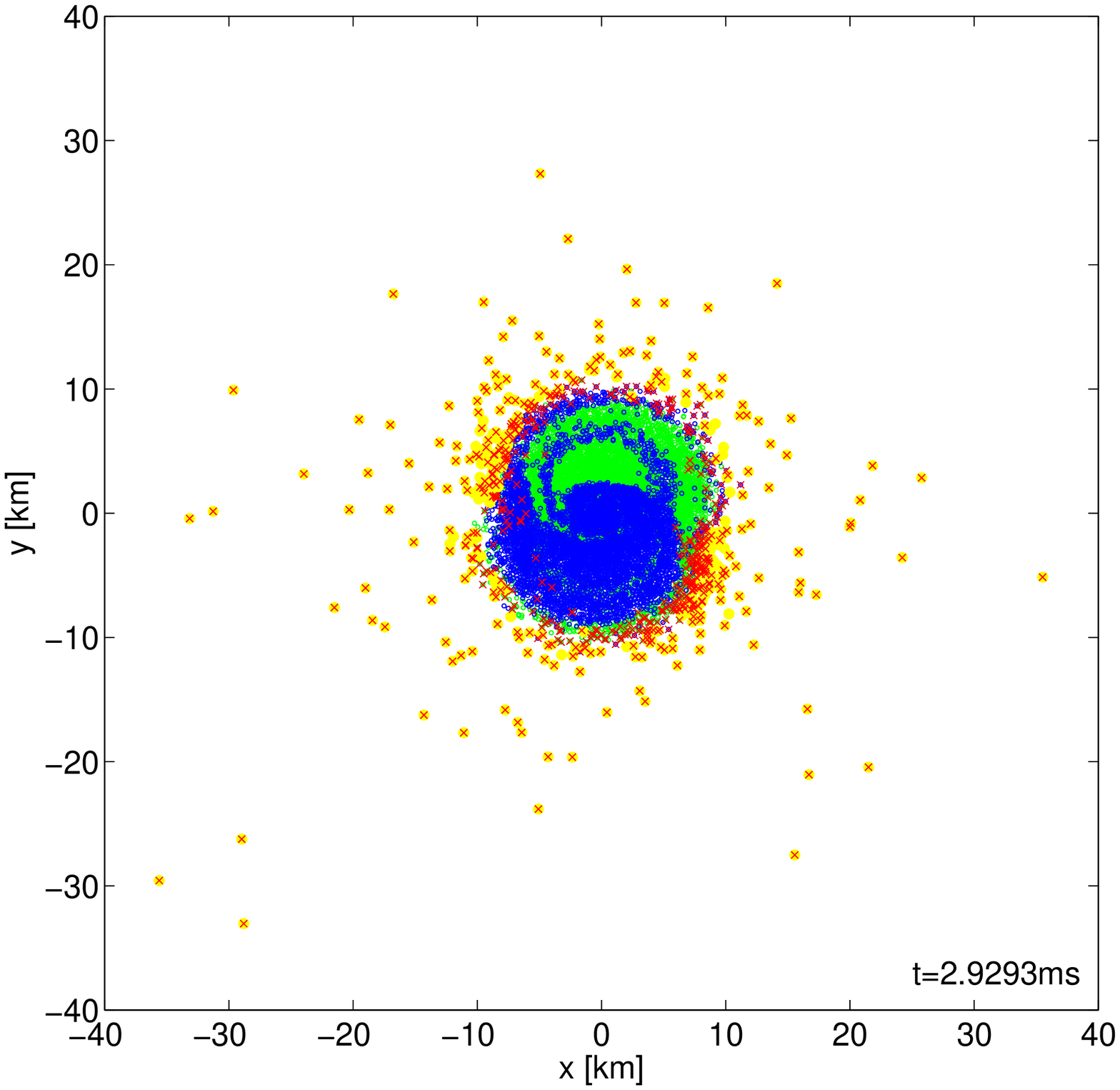}
	(c)
\end{minipage}
  \begin{minipage}[t]{0.47\linewidth}
   \includegraphics[width=7.5cm]{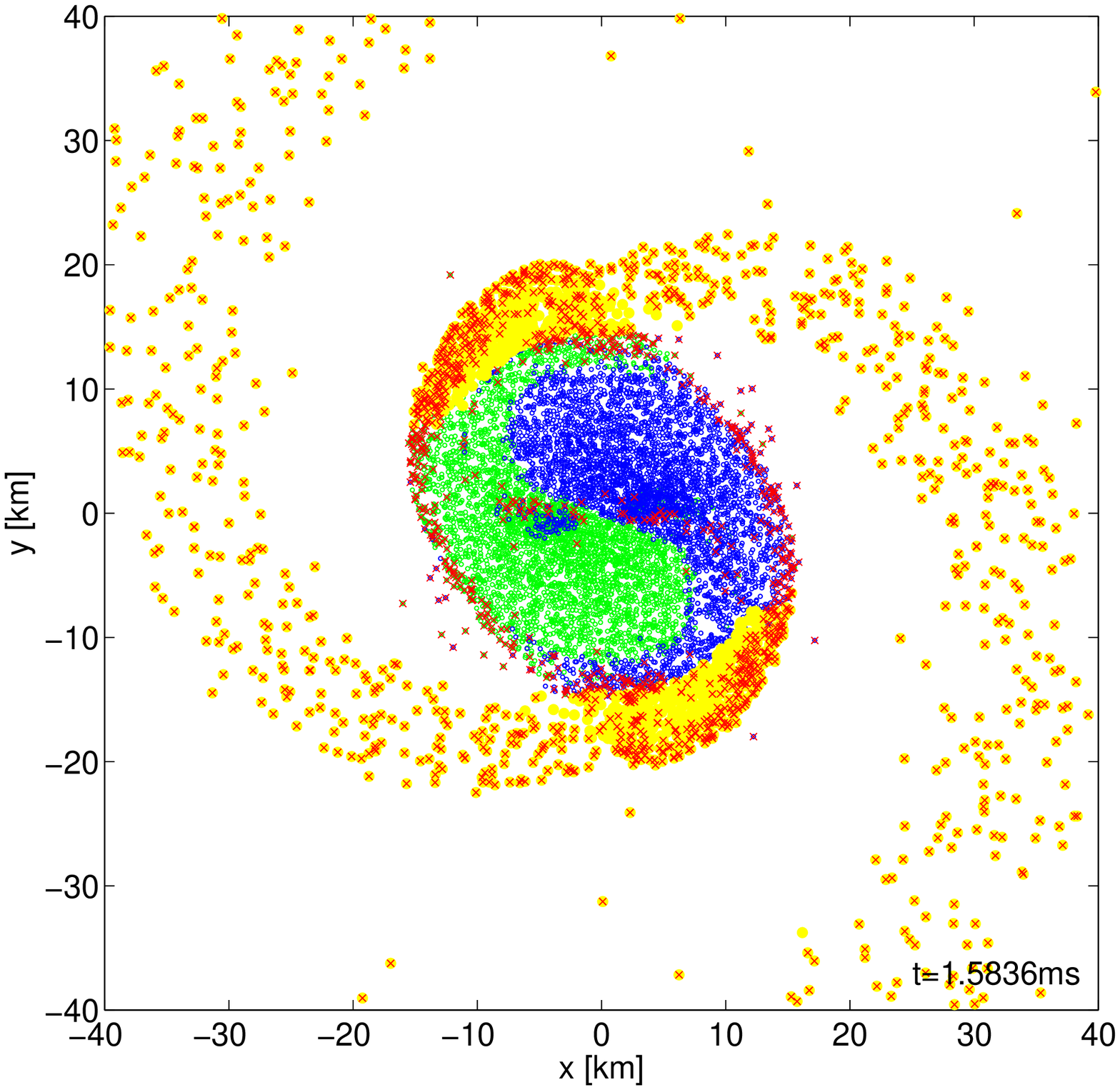}
	(b)

   \includegraphics[width=7.5cm]{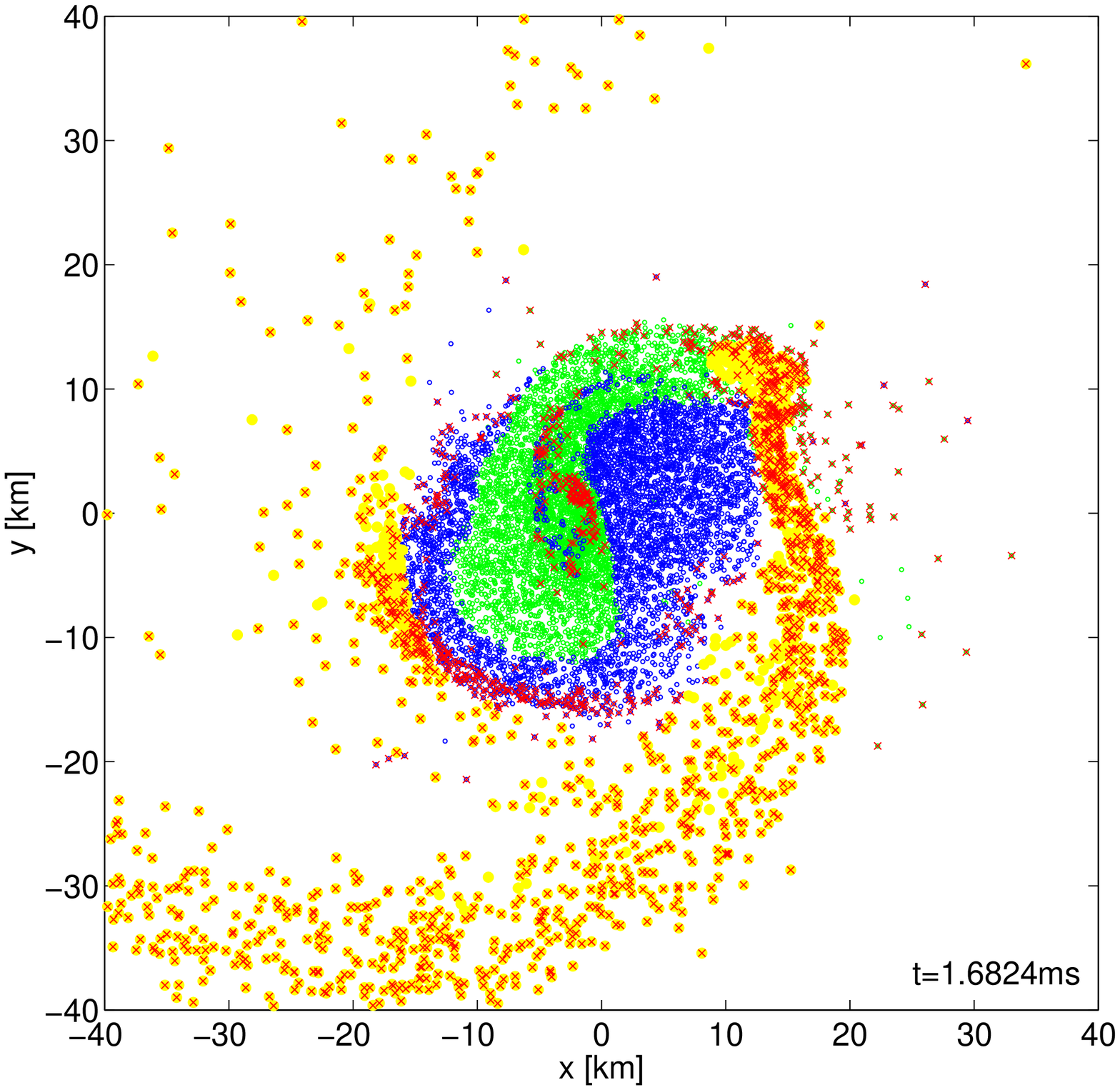}
	(d)
\end{minipage}
\\
\caption{Characteristic snapshots of four representative
models. Plotted is every 10th SPH particle in a slice around the equatorial
plane. The matter of the two NSs is color-coded by green and blue
particles. Particles ending up in the torus are plotted in red while
the yellow circles represent particles which currently have an angular
momentum larger than the angular momentum of the last stable orbit of
the merger remnant at the end of the simulation. (By definition, the
yellow and red particles coincide at the end of the run.). Panel (a)
shows the development of secondary spiral arms in model S1414. Panel (b) shows the large,
primary spiral arms in model S1414co. Panel
(c) shows the post-merger evolution of model L1414. The merger remnant
in much more compact and no secondary spiral arms are forming. As a
consequence, the torus is much smaller and is developing on a longer
timescale. In panel (d), we see the formation of the large primary
spiral arm in model S1216}
\label{fig:particles}
\end{figure}

In models with a mass ratio considerably different
from unity, e.g. in models S107193, S1216 or S1317, the less massive but
larger star is tidally elongated and disrupted to form a long spiral
arm which is largely accreted onto its more massive merger
partner. This leads to a non-axisymmetric, differentially rotating central object and a
thick massive torus around it. The core of the central remnant
forms out of the more massive NS while the accreted material from
the less massive partner is wound up to end in the outer shells of the
remnant. The material sitting in the spiral arm tip receives enough
angular momentum due to a 'sling-shot-effect' to escape accretion and
to contribute to the main part of the future torus. This carries a
significant amount of the angular momentum from the remnant out into the
torus (see Fig. \ref{fig:particles}).
On the other hand, in case of a mass ratio near unity, as eg. in S1414
or S138142 the two partners smoothly plunge together without
any disruption and formation of a primary spiral arm. The two
original NS cores are kept mostly intact during merging and form
together a rotating twin-core, bar-like structure at the center of the merger remnant. This
rotating bar drives the formation of secondary spiral arms at the
remnant surface after the dynamical
merger phase and leads to ejection of material into a small torus. Note that primary spiral arms and large torus masses may
still form in these systems, if the initial NS spin setup is favourable, e.g. in case of corotating NSs.\\
For fixed EoS and NS spin, we find, that the torus mass depends
roughly linear on the mass ratio for $q$-values larger that about
$0.8$, whereas below that point, the torus mass saturates at about
$M_\mathrm{torus}\simeq 0.25M_\odot$. In total, the torus masses range from
about $0.05M_\odot$ at $q=1$ to $0.25M_\odot$ at $q=0.55$ (see
Fig. \ref{fig:discmasses}, panel (b)).

The NS spin setup mainly determines the total amount of angular momentum
in the system and thus directly influences the large-scale postmerger 
evolution, the amount of material carried into the torus and the
spin parameter of the merger remnant. The latter particularly influences
the torus mass via our definiton of the torus. If the NS
are initially corotating, the increase in total angular momentum is about
8\% compared to the irrotating case. Similarly, counterrotating NSs
decrease the total angular momentum by the same amount. 

We find, for fixed EoS (Shen-EoS) and mass ratio ($q=1$), that the
torus mass is indeed mainly determined by the amount of total angular
momentum available in the system whereas details in the merger
dynamics are of minor importance. Models with a different
spin setup, but a similar total angular momentum yield comparable
torus masses. This is the case for models S1414co (corotating) and
S1414t2 (Tilted spins summing up to a total spin in z-direction), and,
to a smaller extent, for models S1414 (no spins) and S1414o (oppositely oriented
spins). This suggests that the torus mass dependence on the NS spin setup can
be approximately reduced to a single parameter, the total angular momentum in the
system (see Fig. \ref{fig:discmasses}, panel (a)). In total, we find torus mass
values between 0.04$M_\odot$ for counterrotating NSs and 0.3$M_\odot$
for corotating NSs.

The EoS influences the merger dynamics and postmerger outcome in
different ways. Initially more compact NSs, as obtained with a soft or
compressible EoS, lead to a longer inspiral phase because the tidal
instability sets in at a smaller orbital separation. Therefore, more
angular momentum is carried away by gravitational wave radiation and the angular momentum in the
post-merger system is reduced which favours a smaller torus. On the other hand,
the angular momentum transport through non-axisymmetric motion, as
e.g. a rotating bar-like core, or through numerical viscosity is found
to be more efficient if the remnant is more compact and more
rapidly rotating. This in turn favours a larger torus.
In model LS1414 (see Fig. \ref{fig:discmasses}, panel (c)), about
twice as much of the angular momentum is radiated away (~15\% compared
to ~8\% in model S1414) and the torus mass shortly (~3ms) after merger
is about half as large (~0.03$M_\odot$ versus ~$0.06M_\odot$). During the following
post-merger evolution the torus grows continously. The ideal gas model
P1315 leads almost to the same torus mass as the Shen-EoS model
S1315. This suggests that the torus mass is determined by the supranuclear part of the EoS.

\section{Implications from observations of short GRBs}

In this section we discuss the implications to the postmerger system
properties from the recent observations and locations of six short GRBs by the Swift and HETE
satellites. Several observational results suggest, that at least part
of the short GRBs may come from binary NS mergers. Short GRBs seem to
be partially associated with elliptical galaxies where star formation has
stopped long ago. They also seem not to be associated with
supernovae. This is consistent with the long inspiral evolution of binary NS systems prior to merger (see e.g. \cite{levan2006}).

From the measured isotropic equivalent energy $E_{\gamma, iso}$, we can
now estimate, within the merger model, the mass which is accreted onto the central BH:
$$
E_\mathrm{\gamma,iso}=f_1 f_2 f_3 f_4 f_\Omega^-1 M_\mathrm{accreted}
$$
Here, $f_1$ is the efficiency at which the accreted rest mass energy
is converted to neutrino emission, $f_2$ is the conversion efficiency
of neutrino-antineutrino annilihation to $e^\pm$ pairs, $f_3$
is the fraction of the $e^\pm$-photon fireball energy which drives the
ultrarelativistic outflow with Lorentz factors $\Gamma > 100$ as
required by GRBs, $f_4$ is the fraction of the energy in
ultrarelativistic jet matter which can be converted to gamma radiation
 in dissipative processes that occur in internal shocks in the jet and $f_\Omega
= 2\Omega_{\mathrm{jet}}/(4\pi) = 1 - \cos\theta_{\mathrm{jet}}$
denotes the jet collimation factor defined as the fraction of the sky
covered by the two polar jets (with semi-opening angles
$\theta_{\mathrm{jet}}$ and solid angles $\Omega_{\mathrm{jet}}$).

 ``Typical'' values from merger and
accretion simulations are: $f_1\sim 0.05$ \cite{lee2005b,
setiawan2004}, $f_2\sim 0.001\,...\,0.01$ \cite{ruffert1999, setiawan2004}, $f_3\sim 0.1$, 
$f_\Omega\sim 0.01\,...\,0.05$ \citep{aloy2005} and $f_4\lesssim 0.2$ \citep[][and references therein]{daigne1998,kobayashi2001,Guetta2001}. To estimate the
accreted mass, we use the set 
\begin{equation}
\label{fvalues}
(f_1,\,f_2,\,f_3,\,f_\Omega,\,f_4) = (0.1,\,0.01,\,0.1,\,0.01,\,0.1)
\end{equation}

to obtain the values in Tab. \ref{tab:grbs}. In addition, we can calculate
the average accretion rate $\dot M_\mathrm{acc}=M_\mathrm{acc}/t_\mathrm{acc}\gtrsim
M_\mathrm{acc}/t_\gamma $ assuming that the observed GRB duration $t_\gamma$ is an
upper limit to the accretion time $t_\mathrm{acc}$ \citep{aloy2005}. Despite
of the crude assumption in (\ref{fvalues}), we find in nearly all cases
values for the torus mass in the range obtained from our theoretical
models. In addition, the estimated mass accretion rates lie in the
ballpark of the results from postmerger accretion disk simulations
\citep{ruffert1999,lee2002,setiawan2004,lee2005b}. Note however that
all the f-factors have large uncertainties and are likely not to be
constant but to depend on the torus mass. For example, the neutrino
annilihation efficiencies sensitively depend on the temperature which in
turn depends on the fluid density and thus on the torus
mass. Therefore, the large values for the torus mass and the accretion
rate from the very luminous burst GRB051221 are not astonishing but
suggest different values for the f-factors or, alternatively, a NS+BH
merger as a source. Newtonian simulations of NS+BH mergers
\citep{kluzniak1998,portegies1998,lee1999,janka1999,rosswog2004,
rosswog2005, davies2005} suggest that, for certain mass ratios and
EoSs, large torus masses in the range of 0.1\,...\,1$M_\odot$ may be formed.

\begin{table}
\caption{Estimated accreted masses, $M_{\mathrm{acc}}$, and lower
bounds for the average mass accretion rates.
The quantity $E_{\gamma,{\mathrm{iso}}}$ is the isotropic-equivalent
$\gamma$-ray burst energy (corrected for the cosmological redshift $z$ of
the burst), and $t_\gamma$ is the GRB duration at the source, computed
as $t_\gamma = T_{90}/(1+z)$ from the measured 90\%-inclusive interval of
high-energy emission. The observational data were taken from
\citet{fox2005} and \citet{soderberg2006}. }
\vspace{0.5cm}
\label{tab:grbs}
\begin{tabular}{l|c|l|c|c|l}
GRB & $z$ & $t_\gamma $ & $E_{\gamma,{\mathrm{iso}}}$
    & $M_{\mathrm{acc}}$ & $\dot M_{\mathrm{acc}}$ \\
\hline
Unit & $\phantom{\displaystyle{\frac{M_\odot^2}{M_\odot}}}$  & sec. & ergs &
$M_\odot$ & $M_\odot/\mathrm{sec.}$ \\
\hline
050509b & 0.225 & 0.033 & $4.5\times 10^{48}$ & $2.5\times 10^{-3}$ & 0.08 \\
050709  & 0.160 & 0.060 & $6.9\times 10^{49}$ & $3.8\times 10^{-2}$ & 0.6  \\
050724  & 0.258 & 2.4   & $4.0\times 10^{50}$ & $2.2\times 10^{-1}$ & 0.09 \\
050813  & 0.722 & 0.35  & $6.5\times 10^{50}$ & $3.6\times 10^{-1}$ & 1.0  \\
051221 & 0.5459 & 0.16  & $2.4\times 10^{51}$ & 1.3 & 8.13 
\end{tabular}

\end{table}

\bibliographystyle{aipproc}   

\bibliography{../discpaper/biblio.bib}

\end{document}